\documentclass[11pt,aps,shownopacs,superscriptaddress,nofootinbib,preprint,axodraw]{revtex4}
\usepackage{graphics}
\usepackage{graphicx}
\usepackage{epsfig}
\usepackage{psfrag}
\usepackage{color}
\usepackage{axodraw}

\usepackage{amsfonts}
\usepackage{amssymb}

\newcommand*{\be}{\begin{equation}}
\newcommand*{\ee}{\end{equation}}
\newcommand*{\bea}{\begin{eqnarray}}
\newcommand*{\eea}{\end{eqnarray}}

\providecommand*{\ler}{\stackrel{\scriptstyle <}{\scriptstyle \sim}}
\providecommand*{\ger}{\stackrel{\scriptstyle >}{\scriptstyle \sim}}

\newcommand{\comment}[1]{}

\newcommand{\cref}[1]{Chapter~\ref{c.#1}}

\def\beq{\begin{equation}}  
\def\eeq{\end{equation}}  
\def\bea{\begin{eqnarray}}   
\def\eea{\end{eqnarray}}   
\def\ba{\begin{array}}   
\def\ea{\end{array}}    
\def\bi{\begin{itemize}}   
\def\ei{\end{itemize}}   
\def\be{\begin{enumerate}}   
\def\ee{\end{enumerate}}   
\def\bc{\begin{center}} 
\def\ec{\end{center}} 
\def\bt{\begin{table}}  
\def\et{\end{table}}    
\def\btb{\begin{tabular}} 
\def\etb{\end{tabular}}

\def\msusy{M_{\rm SUSY}} 
\def\mgut{\, M_{\rm GUT}}

\def\pa{\partial}

\def\asusy{a^{\rm SUSY}_\mu} 
 
 \def\relic{\Omega_{DM 
}} 
\def\lsim{\raise0.3ex\hbox{$\;<$\kern-0.75em\raise-1.1ex\hbox{$\sim\;$}}} 
\def\gsim{\raise0.3ex\hbox{$\;>$\kern-0.75em\raise-1.1ex\hbox{$\sim\;$}}}

\begin{document}

\rightline{LPT--Orsay 07/13}
\rightline{IISc/CHEP/2007/05}
\rightline{April 2007}    

\title{SUSY-GUTs, SUSY-Seesaw and the Neutralino Dark Matter} 
\author{L. Calibbi}
\email{lorenzo.calibbi@uv.es}
\affiliation{Departament de F\'{\i}sica Te\`orica, Universitat de 
Val\`encia-CSIC, E-46100, Burjassot, Spain.}
\author{Y. Mambrini}
\email{yann.mambrini@th.u-psud.fr}
\affiliation{Laboratoire de Physique Th\'eorique,  
Universit\'e Paris-Sud, F-91405 Orsay, France}
\author{S. K. Vempati}
\email{vempati@cts.iisc.ernet.in}
\affiliation{Centre for High Energy Physics, Indian Institute of
Science, Bangalore 560 012, India}
\begin{abstract}
We will consider a SUSY-$SU(5)$ with one right-handed neutrino with 
a large top like Yukawa coupling. Assuming universal soft masses at 
high scale we compute the low-energy spectrum and subsequently
the neutralino LSP relic density taking also into consideration 
$SU(5)$ as well as the see-saw running effects above the gauge coupling 
unification scale. We found that there exists no viable region in 
parameter space for $\tan\beta~\ler ~35$.  The $\tilde{\tau}$ 
coannihilation process starts becoming efficient for $\tan\beta~\ger~35-40$. 
However, this process is significantly constrained
by the limited range in which the stau is lighter than the neutralino.
In fact, for a given $\tan\beta$ we find that there exists an 
upper bound on the lightest neutralino mass ($M_{\chi_1^0}$) in this
region.
The A-pole funnel region appears at very large $\tan\beta~\simeq 45-50$, while
the focus-point region does not make an appearance till large 
($m_0,M_{1/2}$), namely a few TeV. Large $A_0$ terms at high scale can 
lead to extended regions consistent with WMAP constraints and remove 
the upper bounds in the stau coannihilation regions. 
\end{abstract} 
\maketitle

\section{Introduction }
The presence of a natural Dark Matter (DM) candidate has been one of the 
hallmarks of low energy supersymmetry models (SUSY). Within the Minimal
Supersymmetric Standard Model (MSSM),
imposition of R-parity conservation would lead the lightest SUSY 
particle (LSP) to be stable and preferably neutral and a color singlet,
a perfect candidate to explain the Dark Matter (DM) relic density in terms
of a WIMP (Weakly Interacting Massive Particle) \cite{kamionkowski}.
With the advent of precision cosmology, pinnacle of it being the recent 
results from the WMAP experiment \cite{wmap}, the dark matter 
relic density is now known with high accuracy \cite{pdg}. Together 
with availability of state-of-the-art numerical tools it is now 
possible to compute the relic density within
a given SUSY model accurately up to few percent level. 

However, it has been known that within most SUSY models, 
the relic density computed is either too large or too small;
it is required that certain precise or \textit{critical} 
relations between the various soft SUSY breaking parameters 
to exist in order the neutralino relic density comes out 
consistent with the measurements. The accuracy with which these 
relations need to be satisfied typically leaves out very tiny special 
regions in the parameter space where neutralino relic density is 
satisfactory; the tiny regions can be traced back to existence 
of one of the relations between the parameters \cite{welltempered}. 

In the simplest models of gravity mediation, \textit{i.e,} 
minimal supergravity (mSUGRA), which will be the focus of our work, the soft 
terms have `universal' boundary conditions at a high energy scale,
such as the Grand Unification scale, $\mgut$. 
The lightest supersymmetric particle (LSP) is the lightest neutralino 
($\tilde{\chi}_1^0$) in most of the parameter space.
There exists three regions where viable DM relic density is possible
after taking into account the existing low energy direct and indirect 
constraints on the supersymmetric spectrum including the recent 
constraints from LEP measurements \cite{recentworks}. These are the 
(i) The \textit{stau} coannihilation region, (ii) 
The A-pole funnel region and (iii) The focus-point or the hyperbolic 
branch region. In the region (i), where the lightest stau ($\tilde{\tau}_1$)
mass is very close to the LSP mass \cite{griest}, significant enhancement 
in the annihilation cross-section happens due to the coannihilation 
between the lightest neutralino and the $\tilde{\tau}_1$ \cite{staucoann1,
staucoann2}
(ii) In the so-called funnel region, resonant enhancement of the annihilation 
cross-section takes place in the process 
($\tilde{\chi}^0_1 \tilde{\chi}^0_1~ \to~ f \bar{f}$) 
through the intermediate state of ($h,A,H$), where $h,H,A$ are represent
light neutral, heavy neutral and the pseudo-scalar higgs 
respectively \cite{polefunnel}. 
And finally the focus-point region (iii) is the region with large universal
soft parameters which however leads to a small weak-scale value of $\mu$
(the Higgs bilinear coupling in the superpotential), 
raising the Higgsino component in the LSP, and thus leading to significant 
enhancement in the annihilation cross-section \cite{focuspoint}. 

\begin{figure}[t]
\begin{center}
\includegraphics[width=0.90\textwidth]{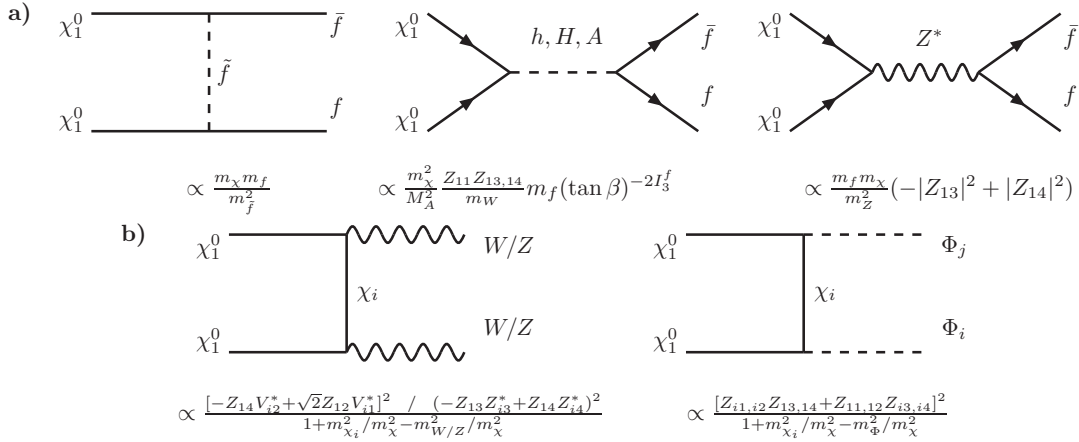}
\caption{Feynman diagrams for LSP neutralino annihilation into a fermion pair
(a) and into massive gauge bosons and Higgs bosons (b). The relevant parts of 
the amplitudes are shown explicitly. $V$ and $Z$ are the chargino and 
neutralino mixing matrices.} 
\label{fig:feynmandetail} 
\end{center}
\end{figure}

A crucial aspect of this arrangement of things where SUSY DM would require
some critical relations to be satisfied between various parameters is that
these relations would not be stable under slightest modifications to the
SUSY soft masses. These modifications are natural when we consider 
extensions of the minimal supersymmetric standard model (MSSM), 
where MSSM  is embedded into either a SUSY Grand Unified Theory (SUSY-GUT) 
and/or augmented with a see-saw mechanism. Additional radiative effects 
present in these models are sufficient to significantly modify the spectrum 
at the weak scale. More importantly, these effects can destabilize the 
special regions in the parameter space where these relations hold or 
sometimes these modifications can be so large that there might be no parameter 
space left where these special relations can hold. This is precisely the 
issue we wish to address in our present work. 

We will consider a SUSY-$SU(5)$ with one right-handed (RH) neutrino 
($SU(5)_{RN}$) with a large
top like Yukawa coupling\footnote{We need at least one more RH 
neutrino and a corresponding Yukawa coupling to fit the neutrino data. 
This Yukawa coupling could be much smaller than the top Yukawa and thus
its effects negligible to the discussion pertaining here.}. Such a model
is motivated from the requirement of non-zero neutrino masses and can have
UV completion in $SO(10)$ based models. Assuming universal soft masses at 
a scale $M_X\simeq 10^{17}$ GeV, we compute the low-energy spectrum 
taking also into consideration $SU(5)$ as well as the see-saw running effects 
above the gauge coupling unification scale ($\mgut\simeq~ 10^{16} {\rm GeV}$). 
Adding the effects of the further running, down to the weak scale, 
we compute the relic density with the weak scale spectrum. 

Firstly, we found that there exists no viable region in parameter 
space for $\tan\beta~\ler ~35$ for neutralino dark matter. Parameter 
space points consistent with WMAP results start appearing with 
$\tan\beta~\ger~35$ and increase with increasing $\tan\beta$. 
However, this process is highly constrained by the limited range 
in which the stau is lighter than the neutralino. In fact, we find 
that the coannihilation region is significantly modified with respect to
the MSSM and it has the shape of a \textit{trunk}, 
in the ($m_0$, $M_{1/2}$) plane\footnote{$m_0$ and $M_{1/2}$ are the 
universal mSUGRA boundary conditions for the scalars and gauginos masses, 
respectively.}, cutting into the $M_{1/2}$ axis. Thus, 
for a given $\tan\beta$, there exists an \textit{upper bound} on the 
lightest neutralino mass ($M_{\chi^0_1}$). The allowed parameter space 
widens up at very large $\tan\beta~\simeq 50$ where the A-pole funnel 
region appears. The focus-point region is very difficult to realise
in this model because electroweak symmetry breaking is much easier
to achieve compared to the standard MSSM. In fact, even going up to 
5 TeV in ($m_0$,$M_{1/2}$) plane, focus-point region does not make
an appearance.

While the above results are particularly true as long as the universal
trilinear coupling $A_0=0$ at the high scale, 
the results can significantly change
with large $-3~m_0~\ler~ A_0~\ler3~m_0$ boundary conditions which 
due to the possibility that some of the masses becoming tachyonic, 
significantly affects the viable parameter space.
For instance, for $A_0 =+3~m_0$, 
the stau coannihilation region branches out
into two regions, a new region other than the \textit{trunk} region 
discussed above. The A-pole funnel also spreads giving rise to a
larger region of the parameter space which is viable under WMAP 
constraints.

The rest of the paper is organized as follows: in the next section, 
we will elaborate our framework along with revising some aspects of
neutralino dark matter. In section 3, we summarize the constraints
we use in our numerical analysis as well as sketch our computing 
procedure. In section 4, we discuss our results and close with a
discussion in our final section. 

\section{$SU(5)_{RN}$ and neutralino DM} 

As mentioned in the introduction, in the following we will consider
a simple SUSY-GUT framework based on $SU(5)$. Further we also add a
RH neutrino with a large top-like Yukawa coupling, which
is natural when this model is incorporated in an $SO(10)$ model~\cite{oscar}.
The superpotential of this framework is given by:
\begin{equation}
\label{su5w}
W_{SU(5)_{RN}} = {1 \over 2}~ h^u_{ii}~ 10_i ~10_i ~5_u
+ h^\nu_{3} ~\bar{5}_3~ 1~ 5_u +
h^d_{ij}~ 10_i ~\bar{5}_j~ \bar{5}_d + {1 \over 2}~M_R~1 1,
\end{equation}
$i,j$ being generation indices and $10$ and $5~(\bar{5})$, representing
the tensorial and fundamental (anti-fundamental) representations of 
$SU(5)$ respectively. The subscript $u,d$ denote the up and down type
Higgs fields\footnote{We have written the superpotential in the basis
where up-type quarks are diagonal.}. Two comments are in order before 
we proceed further. 

Firstly, to accommodate neutrino masses, we would need at least one more
RH neutrino in addition to the one already present in 
Eq.~(\ref{su5w}). However, in most models, the corresponding Yukawa couplings
are not as large as the top Yukawa coupling\footnote{In fact, if one considers
models based on $SO(10)$, it can be as small as the charm Yukawa, $y_c$.}. 
Thus, the impact of
the second and third neutrino Yukawa couplings even if they are present 
is expected to be minimal on the discussion below. In fact, we have 
explicitly checked this to be the case in models where the neutrino 
Yukawa coincides with the up-type Yukawa matrix. Secondly, keeping 
$SO(10)$ completion in mind, we have chosen the high scale $M_X$ to be 
around $\simeq 10^{17}$ GeV. Another natural
choice would be the reduced Planck scale, where $M_X \simeq 10^{18}$ GeV. 
Nevertheless, with this choice of the scale, we believe
our limits on tan$\beta$ (as well as the neutralino mass) will only 
become stronger. In Sec. \ref{relicsec} we will briefly comment about 
the dependence of our results on the choice of $M_X$. 
For most of the discussion, however, we will just choose
the lower value for $M_X$. 

The soft SUSY-breaking terms are considered to be universal at the scale, $M_X$.
These contain the standard mass-squared terms for the scalar 
particles (the squarks and the sleptons), the bilinear Higgs scalar couplings,
the trilinear scalar couplings as well as the mass terms for the gauginos.
Given that R-parity is conserved, the LSP remains stable. 
In most models of SUSY-breaking like mSUGRA, the LSP is the lightest
neutralino. The neutralinos are the physical superpositions of the
fermionic partners of the neutral electroweak gauge bosons, called bino 
($\tilde{B}^0$) and wino ($\tilde{W}_3^0$), and of the fermionic 
partners of the  neutral Higgs bosons, called Higgsinos 
($\tilde{H}^0_u$, $\tilde{H}_d^0$). We can express the lightest 
neutralino as
\begin{equation}
\tilde{\chi}^0_1 = {Z_{11}} \tilde{B} + {Z_{12}} \tilde{W}_3 +
{Z_{13}} \tilde{H}^0_d + {Z_{14}} \tilde{H}^0_u\ ,
\label{lneu}
\end{equation}
where $Z_{ij}$ are the elements of the neutralino mixing matrix. 
It is commonly defined that $\tilde{\chi}^0_1$  is mostly gaugino-like 
if $P\equiv \vert {Z_{11}} \vert^2 + \vert {Z_{12}}  \vert^2 > 0.9$, 
Higgsino-like if $P<0.1$, and mixed otherwise.
This factor $P$ has a major impact in determining the relic abundance 
of the LSP as it determines the rate of neutralino annihilation cross-section.
In Fig.~\ref{fig:feynmandetail} we show the relevant 
Feynman diagrams contributing to neutralino annihilation. This can
be computed for a given SUSY spectrum, which depends on the SUSY
breaking model. 

In the MSSM, with universal soft terms at $\mgut =2\times 10^{16}$ GeV (CMSSM), 
for most of the parameter space,  the lightest neutralino is mainly bino and,
as a consequence, the annihilation cross-section is small producing too 
large relic abundance.  Nevertheless,  as mentioned in the introduction, 
three corridors exist where the cross-section is enhanced and the WMAP 
bounds can be satisfied. First,  there is the coannihilation branch, 
i.e.  the region where the stau mass  is almost degenerate with that of LSP.
Another corridor corresponds to  the A--pole region.  This occurs in the
parameter space where $4 (m_{\tilde \chi^0_1})^2 \simeq m_A^2 \simeq 
m_{H_1}^2 - m_{H_2}^2 - M_Z^2$, with $m_{H_i}^2$ being the soft Higgs mass 
terms, and the dominant neutralino annihilation 
process is  through the `resonant' s-channel pseudo-scalar Higgs exchange.
Finally, there is a `Higgsino' corridor close to the no electroweak symmetry
breaking (No EWSB) region.  In this region the $\mu$ parameter is much smaller 
than the bino mass and the lightest neutralino has a larger Higgsino component 
(thus annihilating efficiently via $Z$ boson exchange). 

As we have stressed in the introduction, these corridors or special
regions in the parameter space are not stable under modifications 
to the soft spectrum. In this connection, it was already noted, for
e.g. in Refs.~\cite{nonu}, that the annihilation cross-section can 
change significantly  if the soft terms determined at the unification 
scale, $\mgut$, are non--universal. Similarly, Ref. \cite{profumo} 
assumed no-scale boundary conditions at $M_X = M_{\rm Planck}$, 
which would result in non-universal boundary conditions at 
$\mgut$\footnote{Our work is closest in spirit to this work and could be 
considered a generalization of it.}. In the scheme we are considering, 
which has been elaborated in the beginning of this 
section, something similar happens: the soft terms are indeed 
non-universal at the GUT scale and, as we will discuss in detail 
below, this would have implications for the regions in the 
parameter space where viable relic neutralino density was possible. 

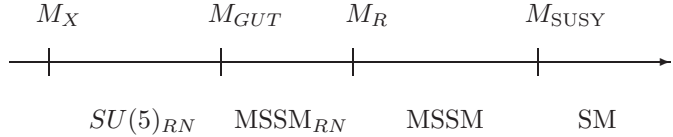
\begin{figure}[t]
{\small
\begin{center}
\begin{picture}(280,65)

\put(0,40){\vector(1,0){250}}

\put(10,55){$M_X$}
\put(15,45){\line(0,-1){10}}
\put(75,55){$M_{GUT}$}
\put(80,45){\line(0,-1){10}}

\put(127,55){$M_{R}$}
\put(130,45){\line(0,-1){10}}

\put(195,55){$M_{\mathrm{SUSY}}$}
\put(200,45){\line(0,-1){10}}

\put(30,15){\small  $SU(5)_{RN}$}
\put(85,15){\small $\mathrm{MSSM}_{RN}$}
\put(150,15){\small MSSM}
\put(215,15){\small SM}
\end{picture}
\end{center}
}
\caption{\label{scales}Schematic
 picture of the energy scales involved in the model.}
\label{integration}
\end{figure}

%
\section{Spectrum and Constraints}
\label{constrainsec}
Before proceeding further in the present section, we will provide
details of our computation of the supersymmetric spectrum in the
framework we have considered and the various constraints we have
used. A schematic diagram describing the integration procedure is presented
in Fig.~\ref{integration}. As inputs at the weak scale, we consider
the Yukawa couplings of the up--type quarks, down--type quarks, charged
leptons and $\tan\beta$. We consider one generation of neutrino with
a mass around $0.05$ eV. We use 1--loop RGEs to run all the Yukawa 
couplings up to the high scale. A more detailed description of the 
procedure can be found in Ref. \cite{calibbi}. 
After several iterations to check 
consistencies between neutrino--scale spectrum and $M_X$, we are now
ready to compute the spectrum from the high scale to the weak scale.
The soft parameters, gauge and Yukawa couplings follow the 
$\mathrm{SU(5)}_{RN}$ RGEs \cite{hisano} between $M_X$ and $\mgut$. We then
compute the evolution of these SUSY parameters using a modified version
of the Fortran package {\tt SUSPECT} \cite{Suspect,Suspect2}
where we have included one RH sneutrino between the GUT
scale and the RH neutrino mass scale, $M_R$. 
The runnings then follows the classical MSSM RGEs after
$M_R$ where the RH neutrino sector decouples.
The low energy mass spectrum is then calculated using {\tt SUSPECT}.
The evaluation of the $b \rightarrow s \gamma$ branching ratio, 
the anomalous moment of the muon and the  relic neutralino density
is carried out using the routines provided by
the program {\tt micrOMEGAs1.3.1} \cite{micromegas}.
The following are the list of constraints we apply on the soft 
spectrum:

\vskip 1cm
\noindent
\textit{(i) Electroweak symmetry breaking :}

\noindent
Minimizing the Higgs potential in the MSSM leads to the standard (tree-level)
relation
\begin{equation}
\mu^2 = 
\frac{-m_{H_2}^2 {\mathrm {tan^2}}\beta + m_{H_1}^2}
{{\mathrm{tan^2}}\beta -1} 
-\frac{1}{2}M_Z^2 \, ,
\label{electroweak}
\end{equation}
This minimization condition is imposed at the scale 
$M_{\rm SUSY}\equiv \sqrt{m_{\tilde t_1} m_{\tilde t_2}}$, 
where $m_{\tilde t_i}$ are the stop masses.
Eq.~(\ref{electroweak}) can be  approximated in most cases by  
\beq
\label{e.ewa}
\mu^2 \simeq - m_{H_2}^2 - \frac{1}{2} M_Z^2
\eeq 
When the right-hand side is negative, electroweak breaking cannot occur. 
The Higgs mass parameter  $m_{H_2}^2$ is positive at the GUT scale, but
decreases with decreasing scale down to $M_{SUSY}$, through the contributions
it receives from RG running at the scale $\tilde \mu$; in presence of 
RH neutrinos, the RG equation for $m_{H_2}^2$ reads: 
\begin{equation}
 (4\pi)^2 {\pa  m_{H_2}^2 \over \pa \ln (\tilde \mu/M_X)} \simeq 
6 y_t^2 (m_{H_2}^2 + m_{\tilde{U}_3}^2 + m_{\tilde{Q}_3}^2 + A_t^2)
+ 2 y_{\nu}^2 (m_{H_2}^2 + m_{\tilde{N}}^2 + m_{\tilde{L}_3}^2 + A_{\nu}^2)
\label{ynurunning}
\end{equation}
\noindent
where gauge contributions have been neglected. Here $m_{\tilde{Q}_3}^2$ and 
$m_{\tilde{U}_3}^2$ corresponds to the $\tilde{t}_L$ and $\tilde{t}_R$ 
soft masses, while $m_{\tilde{L}_3}^2$, $m_{\tilde{N}}^2$ are the same for 
the left-handed and right-handed sneutrinos respectively; $y_t$ and $y_\nu$
are the top and the neutrino Yukawa couplings, while $A_t$ and $A_\nu$ the 
corresponding soft SUSY-breaking trilinear couplings. 
Typically, in the MSSM the value of $m_{H_2}^2$ at the scale $\msusy$ 
depends mainly on the gluino mass $M_3$ at the GUT scale  
(via its effect of increasing  $m_{\tilde{U}_3}^2$ and  $m_{\tilde{Q}_3}^2$).
However, in our case, the further increasing of 
$m_{\tilde{U}_3}^2$ and  $m_{\tilde{Q}_3}^2$ due to the GUT running between 
$M_X$ and $\mgut$, together with the positive additional contribution
 proportional to $y_\nu^2$, drives $m_{H_2}^2$ toward 
even more negative values. As a consequence, the parameter space region
where $\mu^2 < 0$ (or No EWSB region) is much more reduced, 
and even absent in our case as we will see in the analysis.

\vskip 1cm
\noindent 
\textit{(ii) The mass spectrum constraints:}

\noindent 
We have implemented in our analysis the lower bounds on the masses of SUSY particles and of the lightest Higgs boson.
In the squark and slepton sector we checked for the occurrence of tachyons. 
We applied in our analysis the LEP2 lower bound limit on the mass of the lightest 
chargino  $m_{\tilde{\chi}^+_1} > 103.5$ GeV. 
In the non-tachyonic region, typically, the most constraining is the lightest 
Higgs boson mass constraint. 
In the decoupling limit ($M_A \gg M_Z$, applicable in all our parameter space),
$m_h > 114$ GeV at $3\sigma$. 
This bound is very sensitive to the value of the top mass. We have taken
$m_t=174$ GeV throughout our analysis.

\vskip 1cm
\noindent 
\textit{(iii) The $b \rightarrow s \gamma$ branching ratio:}

\noindent 
One observable where SUSY particle contributions might be large is the radiative
flavor changing decay $b \rightarrow s \gamma$. In the Standard Model this
decay is mediated by loops containing the charge 2/3 quarks and $W-$bosons.
In SUSY theories additional contributions come from loops involving charginos 
and stops, top quarks and charged higgs bosons or, in presence of RG induced 
mixing in the down squarks sector, down squarks and gluinos.   
The measurements of
the inclusive decay $B \rightarrow X_s \gamma$ at CLEO \cite{cleo} and BELLE 
\cite{belle}, leads to restrictive bounds on the 
branching ratio $b\to s\gamma$.  We impose in our analysis 
$2.33\times 10^{-4}\leq BR(b\to s\gamma)\leq 4.15\times 10^{-4}$
 at the 3$\sigma$ level. 
We mostly choose $\mu >0$ enabling cancellations between chargino 
contributions and charged Higgs contributions \cite{Roszkowski:2007fd}. 

Before proceeding further, we elaborate a bit more about the 
flavour processes in our model.  Since we are in a Grand 
Unified theory, one would expect large off-diagonal entries to be
generated in the soft matrices at the GUT scale 
due to RG running effects from $M_X$ to $M_{\rm GUT}$. 
This could have large effects in flavour processes at
weak scale. For example, a large off-diagonal entry in the 23 sector of 
the down squarks would contribute dominantly through the gluino 
interactions to the $ b\to s + \gamma $ rate leading to significant constraints 
on that entry, unless of course, there are accidental destructive 
interferences among the various contributions already mentioned above. 
However, in our case, even if the corresponding mass matrices are 
non-universal at the GUT scale, the entries are 
generically small as they 
are proportional to the CKM entries. Some ambiguity can arise in the way 
we handled the neutrino sector. If all the three right handed neutrinos 
are present and the neutrino Dirac Yukawa matrix carries large 
``left"-mixing then they can generate large off-diagonal entries at 
the $M_R$ scale. In such a case a combination of both leptonic and 
hadronic flavour violating constraints have to be used \cite{lavorone}.
On the other hand, it is known that even small departures from universality at
high energy may have a strong effect on $ b\to s + \gamma $, especially tending
to make the constraint weaker \cite{Okumura:2002wa}. In the present case, 
the constraint will be a bit weaker because RH neutrinos in $SU(5)$ only 
affect the down squarks `RR' sector.  
Moreover, we can avoid these effects at least in our framework by choosing 
small left mixing, though flavour violation in the leptonic sector 
might still be significant even in that case \cite{review, calibbi}. 
We postpone such an analysis for future. 

\vskip 1cm
\noindent
\textit{(iv) The anomalous moment of the muon:}

\noindent
We have also taken into account the SUSY contributions to
the anomalous magnetic moment of the muon, 
$a_{\mu}= (g_\mu -2 )/2$.
We used in our analysis the recent experimental results for the muon
anomalous magnetic moment \cite{g-2}, as well as the most recent
theoretical evaluations of the Standard Model contributions
\cite{newg2}.
 It is found that when $e^+e^-$ data
are used the experimental excess in $(g_\mu-2)$ would constrain a
possible SUSY contribution to be 
$7.1 \times 10^{-10}\ler \asusy \ler 47.1 \times 10^{-10}$ at $2\sigma$ level. 
However when tau data is used a smaller discrepancy with the experimental 
measurement is found.
In order not to exclude the latter possibility, when analyzing the parameter 
space with $\mu > 0$  we will simply plot contours with the relevant value 
$\asusy= 7.1\times 10^{-10}$.

\vskip 1cm
\noindent
\textit{(v) Relic Density:}

\noindent
Our basic  assumption is  that the LSP is stable on cosmological time 
scales. Furthermore we will assume that the LSP abundance is thermal. 
Within such framework the regions of the parameter space that lead to
overproduction of dark matter are excluded. On the other hand, the regions 
that yield LSP abundance below the WMAP limit are not considered as excluded 
(though as less favored), but simply  require non--thermal production  
or a dark matter candidate beyond the soft spectrum. 
The WMAP collaboration gives the 3$\sigma$ narrow limit \cite{wmap} 
\begin{equation}
0.087\lsim\relic h^2\lsim 0.138
\label{eq:WMAP}
\end{equation}
on the dark matter relic abundance. 

\section{$SU(5)_{RN}$ and $\relic$}
\label{relicsec}

Before going to present the numerical results of our analysis, let 
us briefly mention the major changes to the spectrum one could 
expect in our framework and the possible implications on the neutralino
relic density. As we have mentioned in the previous section, we have
assumed that the soft masses are universal at a scale $M_X$ which we 
have chosen to be around $10^{17}$ GeV. As far as the neutralino 
relic density is concerned, two major effects on the soft spectrum
due to the $SU(5)$ running between the scale $M_X$ and $\mgut$ and
subsequently to $M_{R}$ can be given as \cite{calibbi}:

\begin{figure}[t]
\includegraphics[width=0.48\textwidth]{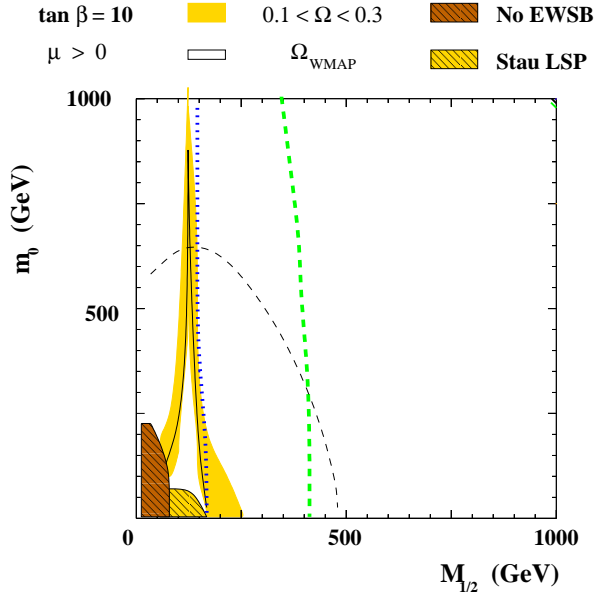}
\caption{The $(m_0,M_{1/2})$ plane with all the low-energy constraints
on the parameter space. While the color code is explained above, the
green dashed line indicates the Higgs mass bound from the LEP, while
the dark dashed line represents the $b\to s+\gamma$ limit. } 
\label{tb10}
\end{figure}

\begin{itemize}

\item The right-handed slepton $\tilde{\tau}_R$ now sits in the 
$10$ and thus it will receive contributions from the full gaugino 
multiplet of the $SU(5)$. At the leading log level, these contributions
are given at the GUT scale as :
\begin{equation}
m_{\tilde{\tau}_R}^2(\mgut)~ \simeq~{144 \over 20 \pi}\alpha_5 M_{1/2}^2  
\ln({M_X \over \mgut}) ~\simeq ~0.25~ M_{1/2}^2 ,
\label{pregut}
\end{equation}
where we have taken the limit $m_0 \to 0$ and $\alpha_5=\frac{g_5^2}{4\pi}$ 
represents the unified gauge coupling at $\mgut$. This large positive
contribution from to the $\tilde{\tau}_R$ make the stau heavier
than the LSP (which is the lightest neutralino) for most of the
parameter space. Let's note here that choosing a larger value for $M_X$ 
(such as $M_{\rm Planck}$) would further increase $m_{\tilde{\tau}_R}^2$,
strengthening the effect described above. 

\item In our framework the neutrino Yukawa coupling taking 
part in the see-saw mechanism as shown in Eq.~(\ref{su5w})  is 
taken to be as large as the top Yukawa coupling. This introduces
an additional top-Yukawa like coupling to the up-type Higgs from 
the scales $M_X$ down to $M_{R}$. Compared to the case of 
MSSM with RH neutrinos, as shown in Eq.~(\ref{ynurunning}), here the length of
running is more and thus more effective radiative electroweak
symmetry breaking takes place\footnote{The value of $M_R$ is uniquely 
fixed ($M_R\approx 6 \times10^{14} {\rm GeV}$), through the see-saw mechanism, 
by our choice of the neutrino Yukawa coupling and light neutrino mass scale. 
Some variation in $M_R$ can come in a complete 
three generation model choosing inverse/normal hierarchy and particular
mixing patterns. Choosing a larger light neutrino mass 
($\mathcal{O}({\rm eV})$) would simply lower $M_R$, thus increasing the RG 
effect on $\mu$.}. 
Moreover, the increasing of the up-squarks soft 
masses, due to the unified gauge sector, contributes to pushing 
$m^2_{H_2}$ down to more negative values than in the CMSSM.
As a consequence, for most of the parameter
space, we find that electroweak symmetry breaking is viable 
unlike in the CMSSM where substantial amount of parameter space
is ruled out. 

\end{itemize}

The above two reasons are sufficient to offset the conditions
which give viable dark matter in CMSSM at low $\tan\beta$. 
As stated in the introduction, we found no viable region giving
the required relic density up to $\tan\beta\ler 35$. In Fig.~\ref{tb10},
we show the non-existence of any region of the parameter space which
gives rise to the correct relic density in our framework of $SU(5)_{RN}$,
for $\tan\beta = 10$ and $A_0=0$. 
All the three branches of CMSSM are not possible here\footnote{Even if we 
have confined ourselves, in the figure, to the 
parameter space below 1 TeV for both $m_0$ and $M_{1/2}$, we have performed 
a scanning up to 5 TeV, without finding viable regions.} as none of 
their corresponding relations could be satisfied within this regime. 

\begin{figure}[t]
\includegraphics[width=0.48\textwidth]{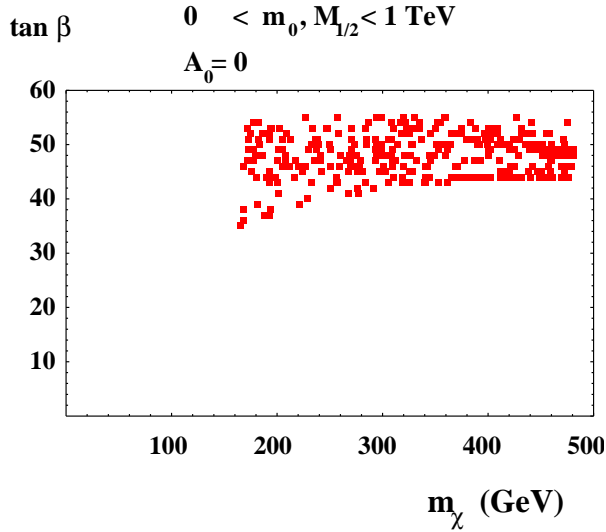}
\caption{Points allowed by experimental and theoretical constrains
after a scan on ($0<m_0,M_{1/2}<1~\mbox{TeV}$) and ($20<\tan\beta<55$)} 
\label{coanntb40}
\end{figure}

To make this statement more concrete, in Fig.~\ref{coanntb40}, we plot
all the points which satisfy the available direct/indirect constraints 
and give viable relic density as a function of $\tan\beta$ and LSP 
mass. From the figure we see that viable DM is only possible for values
of:
\begin{equation}
\mbox{tan}\beta~\ger~34 ~;~~~~~~ m_{\tilde{\chi}^0_1}~\ger 150~ \mbox{GeV}
\label{lowerbounds}
\end{equation}
These are quite strong \textit{lower bounds} on the neutralino mass 
and $\tan\beta$ and will be useful in distinguishing this model compared
to the standard CMSSM parameter space. In the Fig.~\ref{coanntb40},
we see that the small strip of points on the lower left corner gives
a more constrained solution in $m_{\tilde{\chi}^0_1}$ for $\tan\beta~\ler$ 45. 
For larger $\tan\beta$, we see that almost all possible values are allowed
in the lightest neutralino mass. However, at this level, we are unable 
to distinguish
between the actual annihilation processes involved in generating the correct
relic density. A more detailed look shows us the existence of two regions which
we elaborate here.

\subsection{Coannihilation Region}

\begin{figure}[t]
\includegraphics[width=0.48\textwidth]{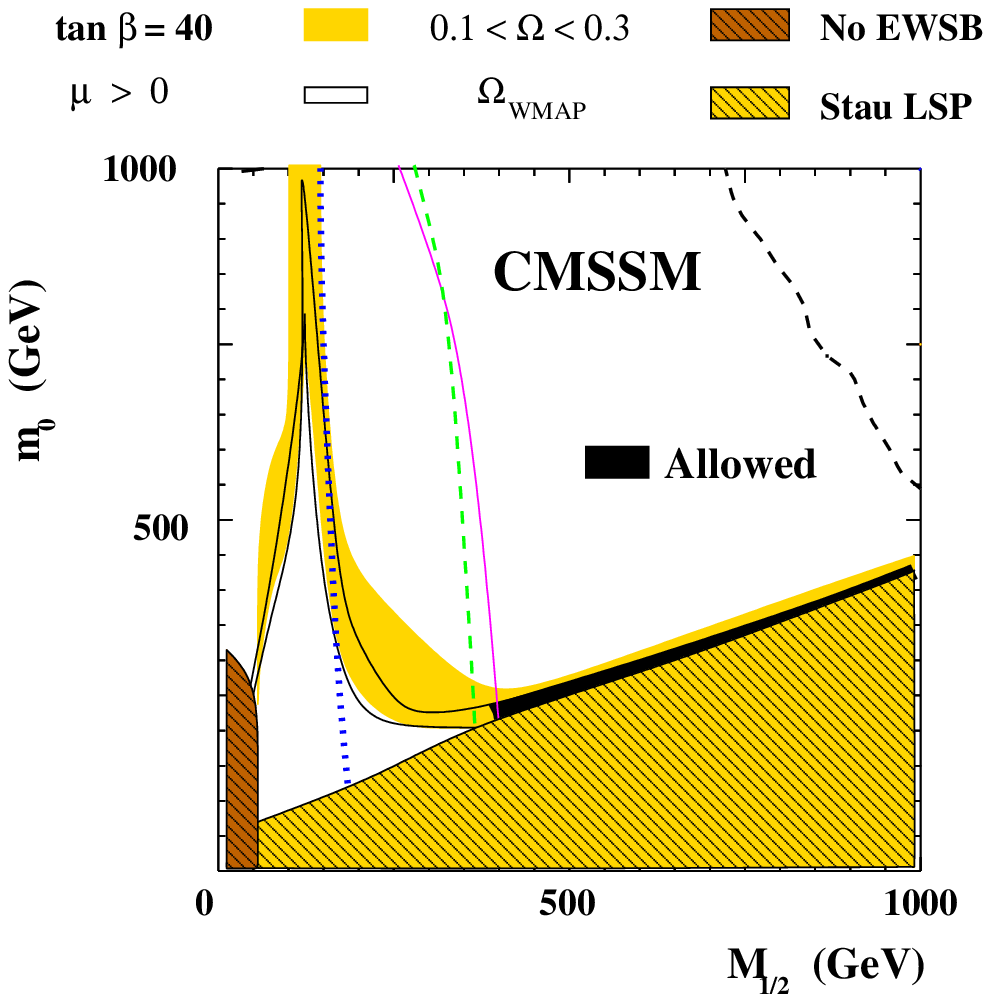}
\includegraphics[width=0.48\textwidth]{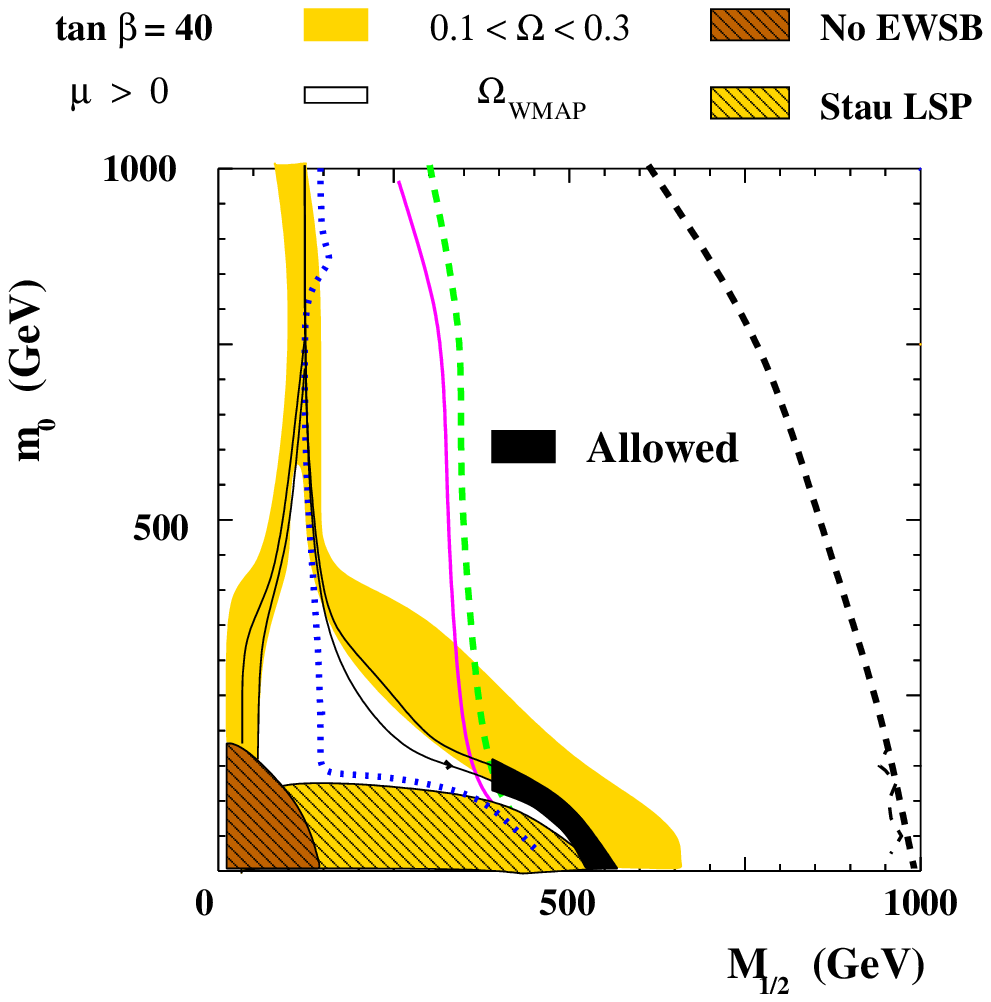}
\caption{We show the coannihilation region for $\tan\beta$=40 for the
case of CMSSM and $SU(5)_{RN}$. The green dashed and the dark dashed
lines represent same as above.} 
\label{tb40}
\end{figure}

As we have discussed previously, the large pre-GUT scale
contributions to the $\tilde{\tau}_R$ in Eq.~(\ref{pregut}) make
the lightest stau heavier than the LSP in most of the parameter
space. However, when $\tan\beta$ becomes large, it is possible
to make the lightest stau $\tilde{\tau}_{1}$ closer
to the LSP mass by making the left-right (LR) mixing 
term $m_\tau(A_{\tau}-\mu\tan\beta)$ large. This is what indeed happens
in our case. Stau coannihilation cross-sections start becoming
effective  for $\tan\beta \ger 34$, leading to relic density 
within the WMAP bound. In Fig.~\ref{tb40} we have plotted,
the WMAP compatible regions for the case of CMSSM as well as
$SU(5)_{RN}$, for $\tan\beta=40$ and $A_0=0$. 

Two features are evident from such a comparison between the
CMSSM and the $SU(5)_{RN}$: (i) the WMAP compatible region
is much smaller in the case of $SU(5)_{RN}$ and (ii)
the shape of the allowed region is quite different in the 
case of $SU(5)_{RN}$ (we will call it {\it trunk} region). 
In fact, the allowed region cuts-off for a value of 
$M_{1/2}~\simeq~520$ GeV in the above plot. This would correspond
to a LSP mass of around $240$ GeV; an LSP of higher mass would
not be able to give rise to the correct stau-neutralino coannihilation
rate. This we would say corresponds to an \textit{upper bound} on 
the LSP mass which could give viable stau coannihilation rate 
such that the relic density is within the WMAP bound. 
The presence of such upper bound in the LSP mass is not peculiar of the
model we are considering here, since similar bounds are well-known also 
within CMSSM \cite{staucoann1, lspbounds1}. 
However, the nature of these upper bounds
is quite different in the two models. In the CMSSM, the reason is that
the coannihilation cross-section decreases at larger SUSY masses 
\cite{staucoann2} and the coannihilation rate becomes too low even if 
the condition $m_{\tilde{\tau}_1}\approx m_{\tilde{\chi}^0_1}$ is satisfied.
On the other hand, in $SU(5)_{RN}$, the region for which 
$m_{\tilde{\tau}_1}\approx m_{\tilde{\chi}^0_1}$ simply disappears for some 
value of $M_{1/2}$, as a consequence of the peculiar shape of the 
$\tilde{\tau}$-LSP region. 
This excludes the possibility of efficient coannihilation even 
for values of $m_0$ and $M_{1/2}$ outside the ``1 TeV box'' 
of Fig.~\ref{tb40} (as we numerically checked),
so that the value $m_{\tilde{\chi}^0_1}\simeq240$ GeV is a real upper 
bound for the case considered ($\tan\beta=40$, $A_0=0$).

The reason for these peculiar features of $SU(5)_{RN}$ can again be traced back 
to the large gaugino contribution to the stau mass above the GUT scale. 
At the weak scale, roughly the stau mass matrix is now given by:
\begin{equation}
\mathcal{M}_{\tilde{\tau}}^2 = \left( \begin{array}{cc}
m_{\tilde{\tau}_{LL}}^2 & m_{\tilde{\tau}_{LR}}^2  \\ 
m_{\tilde{\tau}_{LR}}^2 & m_{\tilde{\tau}_{RR}}^2 \end{array}
\right), 
\label{slepmass}
\end{equation}
where, including the pre-GUT effects,
\bea 
m_{\tilde{\tau}_{RR}}^2~\simeq~(1-\rho)\,m_0^2 + 0.3\,M_{1/2}^2 
\label{mstauR}
\eea
with $\rho$ being a positive coefficient dependent on 
$\tan\beta$ and smaller than one in the present case\footnote{
On the contrary, when $\left|A_0\right| \neq 0$ the parameter $\rho$ could be
driven larger than one by running effects due to the A-term. 
This could cause the arising of tachyonic stau masses, as we will discuss 
in Sec. \ref{subsecA0}.}, and
\bea
m_{\tilde{\tau}_{LR}}^2&=&m_\tau~(A_\tau - \mu \tan \beta)
\nonumber\\ 
&\simeq& - m_\tau~ \mu \tan \beta
\eea
where we explicitly indicate the dominance of the $\tan\beta$ enhanced term.
In first approximation the lightest eigenvalue of 
Eq.~(\ref{slepmass}) is given by:
\bea 
m^2_{\tilde{\tau}_1}&\simeq&  m^2_{\tilde{\tau}_{RR}} - 
m_\tau~ \mu \tan \beta
\nonumber\\
&\simeq& (1-\rho)\,m_0^2 + 0.3\,M_{1/2}^2  - m_\tau~ \mu \tan \beta
\label{lightstau}
\eea
The coannihilation condition requires that $m_{\tilde{\tau}_1}$ 
should be almost degenerate with the mass of the LSP, which is mostly the
Bino mass, $M_1~\simeq~ 0.47\, M_{1/2}$. Taking $m_0 \approx M_{1/2} \approx
\mu \approx \msusy$ and barring the $\tan\beta$ dependence of the coefficient
$\rho$, we find as a result 
$\tan\beta \simeq (1-\rho)(\msusy/m_{\tau})$,
from which, for reasonable values of $\msusy~\ger 200\, {\rm GeV}$ 
and considering that in the present case $0.1\ler\rho\ler0.7$,
we get the lower limit $\tan\beta~\ger~35$\footnote{A longer
running, as in the case $M_X = M_{\rm Planck}$, would increase 
$m^2_{\tilde{\tau}_{RR}}$ in Eq. (\ref{lightstau}). As a consequence an 
even larger value of $\tan\beta$ would be necessary to satisfy to 
coannihilation condition (though the larger $m^2_{\tilde{\tau}_{RR}}$ would
be partially compensated by the simultaneous enhancing of $M_1$).}.  
Solving numerically the system with the parameters varying in a wide region
of the parameter space ($0\le m_0,\,M_{1/2}\le 1\,{\rm TeV}$, 
$5\le \tan\beta\le 50$, $A_0=0$), we find that
that there is no solution for $\tan\beta~\ler~27$, furthermore, for a
given $\tan\beta$, only a restricted region of the parameter space 
in $M_{1/2}$ is allowed. In Fig.~\ref{coanntb40}, this region 
corresponds to the lower strip of points $34 ~\ler\tan\beta~\ler 45$.
As we can see, the limit on $\tan\beta$
is greater after applying the experimental constrains mainly due to 
restrictions coming from the bound on $m_h$.
Overall, the spectrum prefers a neutralino mass of at least of 
$\mathcal{O}(150~\mbox{GeV})$ and further
with an upper bound of about $\mathcal{O}(200~\mbox{GeV})$ if 
stau coannihilation is the dominant process in efficiently reducing
the relic density to the observed levels\footnote{Such a narrow range
can perhaps be useful to distinguish this model compared to standard
CMSSM at colliders.}. 

\begin{figure}[t]
\includegraphics[width=0.48\textwidth]{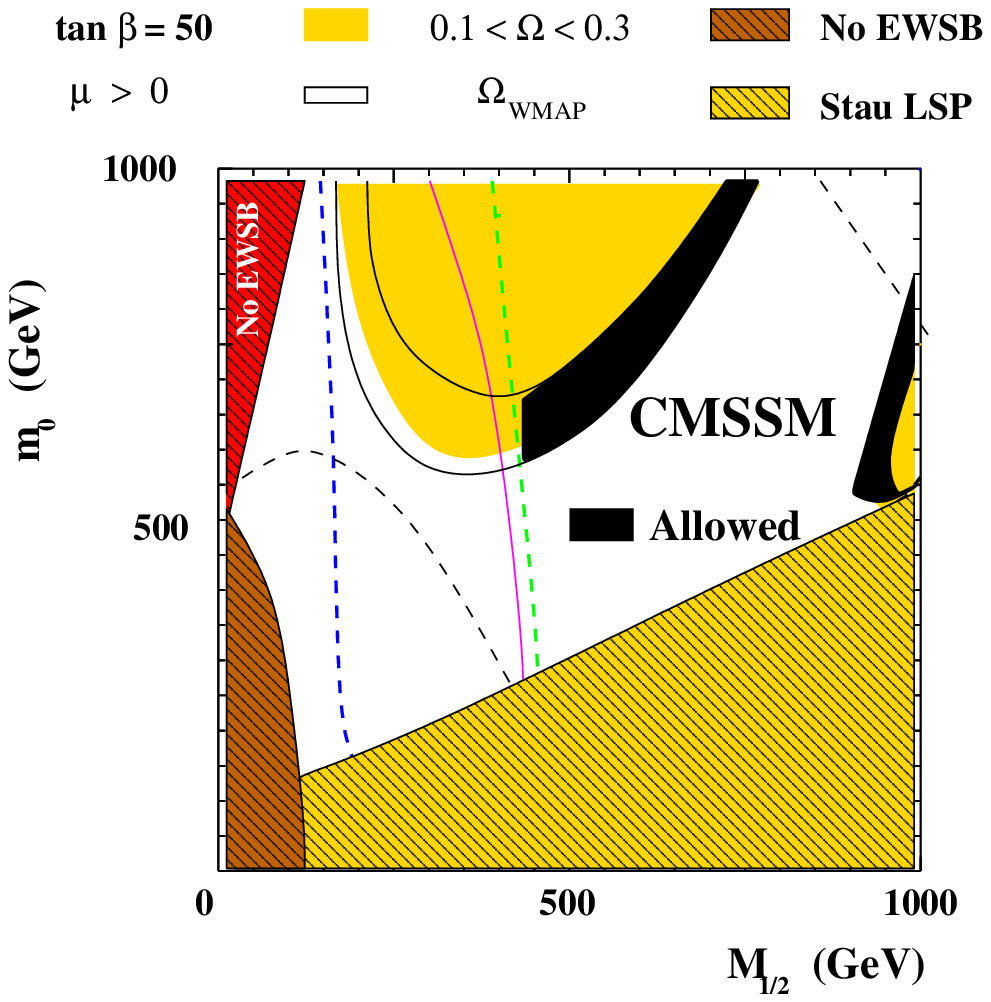}
\includegraphics[width=0.48\textwidth]{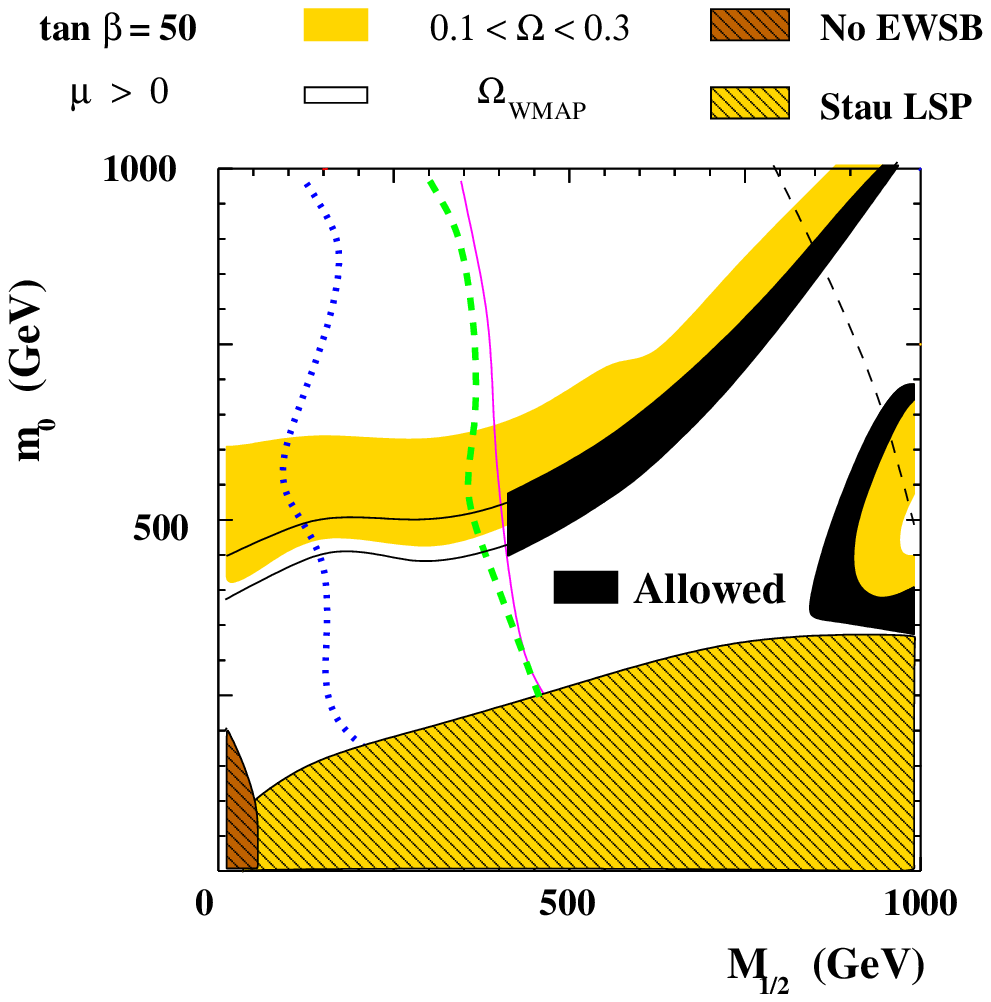}
\caption{Same as above for $\tan\beta=50$.} 
\label{tb50}
\end{figure}

\subsection{A-pole funnel}
Let us now move to the funnel region. As in the CMSSM, this region 
makes its appearance for large $\tan\beta ~\simeq~ 45-50$.
In Fig.~\ref{tb50}, we have plotted the funnel region for $\tan\beta$ = 50, 
$A_0=0$, again in comparison with the CMSSM case. 
We can also see large regions of the parameter space where the lightest 
stau is the LSP and thus there are regions of coannihilation also which 
are fused with the funnel region. The trunk region is no longer present
and neither do the upper bounds.

\subsection{Focus Point region}
As discussed in Sec. \ref{constrainsec}, 
the REWSB is very easy to achieve due to the presence of additional
GUT effects and top like Yukawa contribution from the neutrino Yukawa 
coupling to the up type Higgs between $M_X$ and $M_R$. 
The result is there is no focus-point region present. 
We could not find the focus-point until
$(m_0,M_{1/2}) \simeq 5 {\rm TeV}$. We have not explored further, since
these regions would be far from the reach of LHC.

\subsection{$\left|A_0\right| =3~m_0$}
\label{subsecA0}
\begin{figure}[t]
\includegraphics[width=0.48\textwidth]{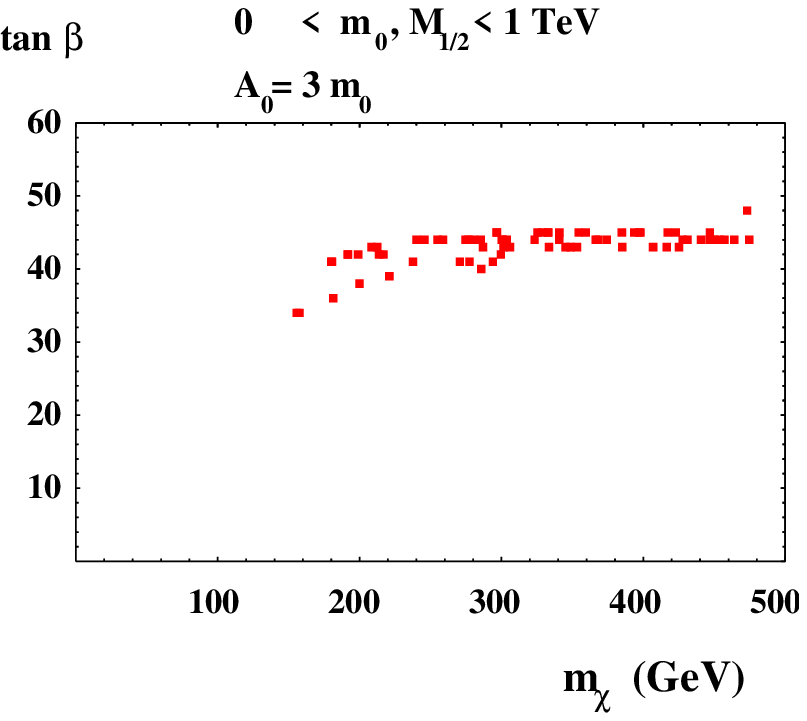}
\includegraphics[width=0.48\textwidth]{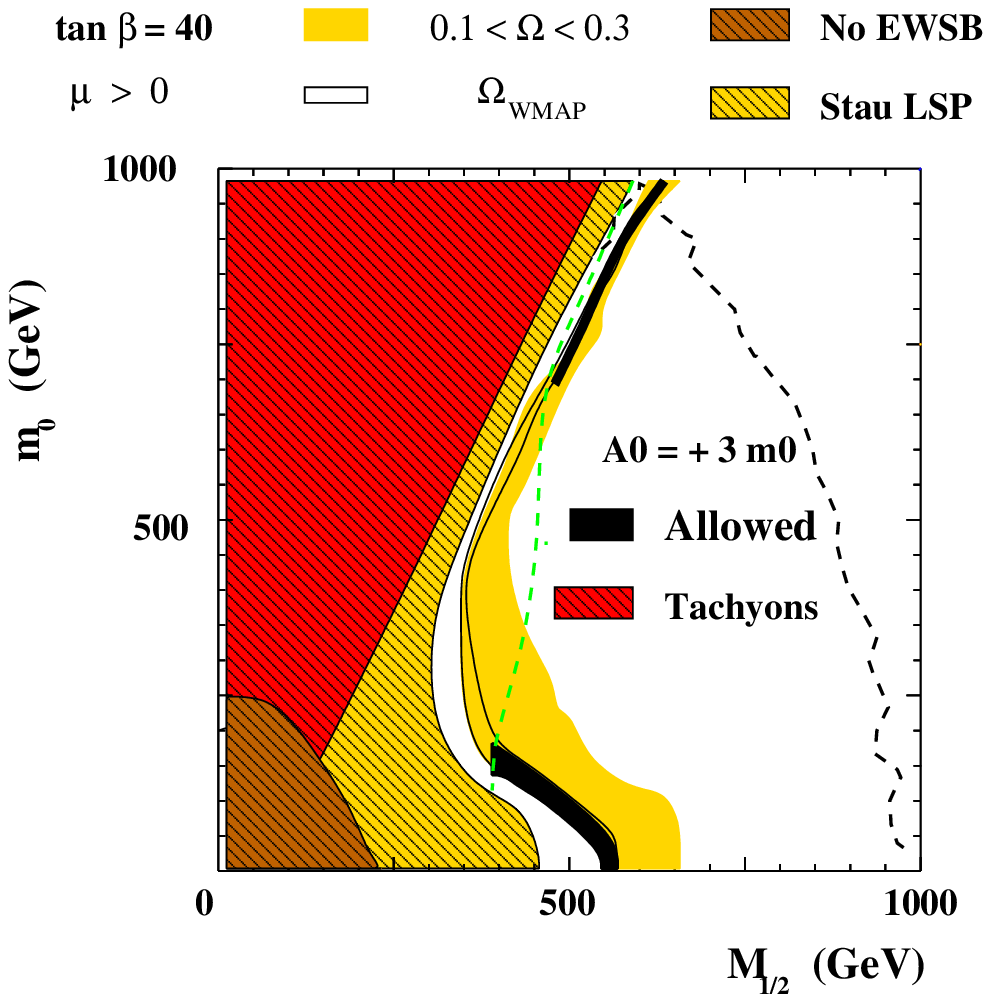}
\caption{$A_0$ = 3~ $m_0$ case. Left panel: points allowed by experimental 
and theoretical constrains
after a scan on ($0<m_0,M_{1/2}<1~\mbox{TeV}$) and ($20<\tan\beta<55$).
Right panel: the $(m_0,M_{1/2})$ plane plot for $\tan\beta=40$; we can see 
the two branches of the stau coannihilation region.} 
\label{a0plus3}
\end{figure}

All the solutions discussed above are for the special case where
$A_0=0$ at the high scale. Given the constrained solutions we have
here specifically for the stau coannihilation process, which can depend
significantly on the non-zero $A_0$ parameter, it is instructive 
to consider the extreme cases of $A_0=-3~m_0$ and $A_0=+3~m_0$. 
For $A_0=-3~m_0$, we found that the occurrence of large regions 
of the ($m_0$, $M_{1/2}$) plane excluded by tachyonic $m^2_{\tilde{\tau}_1}$ 
makes such case really disfavoured, leaving only a small portion of 
the parameter space viable. On the other hand, for $A_0=+3~m_0$
such excluded region results much smaller and localized in the portion
of the plane where $m_0$ is large and $M_{1/2}$ small.
Such behaviour can be explained by looking again at Eq.~(\ref{lightstau}).
If the parameter $\rho$ there, which depends on $Y^2_{\tau}$ and $A^2_{\tau}$,
becomes greater than one because the additional A-term running effect
with respect to the case $A_0=0$, the coefficient in front of $m_0^2$ 
becomes negative. Therefore, values of $m_0^2$ larger than $M_{1/2}^2$
can easily make the lightest slepton tachyonic. 
The effect is enhanced by growing
$\tan\beta$ both because the direct $\tan\beta$ dependence in 
Eq.~(\ref{lightstau}) and because the $\tan\beta$-enhanced 
$Y^2_{\tau}$ and $A^2_{\tau}$ contribution to the value of $\rho$.
The different behaviours for $A_0=\pm 3~m_0$ are due to the renormalization
group running of $A_{\tau}$ itself, which is not independent of the sign
of $A_0$.
In Fig.~\ref{a0plus3}, we plot, for the $A_0=+3~m_0$
boundary condition, the regions which have the
relic density consistent with WMAP measurements. We see from the
left figure that $A_0=+3~m_0$ is very seriously constrained 
in the ($\tan\beta$, $M_{\chi^0_1}$) plane compared to the $A_0=0$ case. 
The appearance of an upper bound ($\approx 45$) on $\tan\beta$ is due to the
increasing of the mentioned tachyonic-$\tilde{\tau}$ region for large values 
of $\tan\beta$, as it appears evident from Eq.~(\ref{lightstau}). 
Moreover, a closer look reveals
a different branch of the coannihilation region in addition to the
trunk region. This is shown in the right panel of Fig.~\ref{a0plus3}. 
We see in addition to the standard trunk branch, a thin strip on the top left
corner, which appears along the $\tilde{\tau}$-LSP 
branch not present for $A_0=0$. Such branch runs along the large 
tachyonic-$\tilde{\tau}$ region which now appears 
due to the non-zero A-term as explained above.

\section{Remarks}
If supersymmetric standard models do really provide a dark matter 
candidate whose relic density constitutes most of the energy budget
of the universe, one would typically expect non-trivial relations
to hold true between various supersymmetric parameters. The point
we tried to address was how stable these relations are under 
modifications to SUSY standard models. SUSY-GUTs with see-saw 
are natural extensions of MSSM in the light of experimental
evidence for non-zero neutrino masses. In the simplest model we
have considered here, $SU(5)_{RN}$, we found
that relic density requirements put severe constraints on the 
range of $\tan\beta$. Only large $\tan\beta~\ger~ 35$ is now
allowed. Such large $\tan\beta$ will also lead to large effects in 
flavour physics \cite{review} which will also be present in this
class of models. 

Of the three possible corridors of viable parameter space present 
in CMSSM only two appear here. The stau coannihilation region 
is both bounded from above and below as long as $A_0 ~=~ 0$.
The A-pole funnel appears for large $\tan\beta$, whereas focus-point
region does not make an appearance in this class of models. 

In the recent times, there has been a large interest in trying
to reconstruct the dark matter relic density from SUSY spectrum
measurements at colliders \cite{peskin,pole}. While LHC alone might
not be able to reconstruct the spectrum, one expects that astrophysical
relic density measurements and LHC data would give a rough handle
on the lightest neutralino mass. The couplings and cross-sections leading to
precision measurements of the SUSY soft spectrum are possible
at the ILC (International Linear Collider) operational in two modes
with $\sqrt{s}= 500$ GeV and 1 TeV. In particular, the use of 
$\tau$ polarization \cite{rohini} should be able to distinguish this model from
CMSSM as long as SUSY spectrum lies in the stau coannihilation
region \cite{ours2}. Here we use the fact that in our model, 
the relative sizes of $m_{\tilde{\tau}_{RR}}^2$, $m_{\tilde{\tau}_{LL}}^2$ and
$m_{\tilde{\tau}_{LR}}^2$ change with respect to the CMSSM and consequently
the large left-right mixing of stau required to get the degeneracy 
with the LSP.
Further, it might be possible
that the trunk region of this model, which has upper and lower bounds in the 
neutralino mass, can be tested from the data at LHC. It is not clear if the 
other regions which can have degenerate solutions with CMSSM can be 
discriminated either at LHC or a combination of LHC and future linear colliders.
Such a question needs further studies. 

Let us conclude by stressing that the peculiar phenomenology of the 
coannihilation we presented here turns out to be a quite general feature
of GUT models. In fact, the crucial point is the increasing of the lightest
stau mass due to effects of the unification gauge group such as
in Eq.~(\ref{pregut}). Similar effects are present every time 
$m_{\tilde{\tau}_{R}}^2$ feels the presence of a unified gauge sector, 
independently of particulars of the GUT structure and of the 
presence of right-handed neutrinos as in our case. Therefore, we claim that 
even in the absence of see-saw or large neutrino Yukawa couplings, 
the phenomenology of stau coannihilation regions would be similar in most 
GUT models.
The presence of large see-saw couplings would effect the radiative 
electroweak symmetry breaking part and thus would have implications
for the focus point regions. In the end our work shows that dark matter
seems to predict large $\tan\beta$ in SUSY-GUTs. 
 
\textbf{Acknowledgements} 
We are highly grateful for discussions with D. P. Roy, X. Tata, R. Godbole 
and G. Polesello which have been highly informative and useful. We also
acknowledge conversations with K. S. Babu, B. Ananthanarayan and P. Paradisi. 
SKV also acknowledges the organizers of the ``From Strings to LHC"
workshop at Goa, Jan, 2007, for the stimulating atmosphere where part 
of the work is done. The work of YM is sponsored by the PAI 
program PICASSO under contract PAI--10825VF. LC acknowledges the ``Angelo
Della Riccia'' foundation, from which his work is supported, and 
the Theoretical Physics Department of the Valencia University for the 
kind hospitality and the stimulating environment.

\end{document}